\documentclass[prb,twocolumn,showpacs,amsmath,amssymb,aps,10pt,citeautoscript]{revtex4-1}

\usepackage{amsmath,amssymb}
\usepackage{graphicx}
\usepackage{dcolumn}
\usepackage{multirow}
\usepackage{braket}
\usepackage{mathtools}
\usepackage{bm}
\usepackage{hyperref}

\newcolumntype{d}[1]{D{.}{.}{#1}}

\hypersetup{%
  colorlinks=true,
  allcolors=blue,
  bookmarksnumbered=true,
}

\newcommand{\rr}{\mathbf{r}}
\newcommand{\RR}{\mathbf{R}}

\newcommand{\kk}{\mathbf{k}}
\newcommand{\nk}{{n\mathbf{k}}}

\renewcommand{\Im}{\operatorname{Im}}

\renewcommand{\Re}{\operatorname{Re}}
\DeclarePairedDelimiter\abs{\lvert}{\rvert}
\begin{document}

\title{Automated construction of maximally localized Wannier functions for bands with nontrivial topology}
\author{Jamal I. Mustafa}
\email{jimustafa@gmail.com}
\author{Sinisa Coh}
\author{Marvin L. Cohen}
\author{Steven G. Louie}
% \email{sglouie@berkeley.edu}
\affiliation{Department of Physics, University of California at
  Berkeley and Materials Sciences Division, Lawrence Berkeley National
  Laboratory, Berkeley, California 94720, USA}

\begin{abstract}
We show that an optimized projection functions method can automatically construct maximally localized Wannier functions even for bands with nontrivial topology.  We demonstrate this method on a tight-binding model of a two-dimensional $\mathbb{Z}_2$ topological insulator, on a three-dimensional strong $\mathbb{Z}_2$ topological insulator, as well as on first-principles density functional theory calculated valence states of Bi$_2$Se$_3$. In all cases, the resulting Wannier functions contain large imaginary components and are more extended than those in the topologically trivial phase.
\end{abstract}

\maketitle

\section{Introduction}\label{sec:intro}

A useful representation of the occupied states in a periodic insulator is the Wannier function. Wannier functions (WFs) provide a localized real-space description of the extended Bloch states.~\cite{MLWFs-Review} In particular, WFs give a chemical picture of the bonding nature of a material, an alternative real-space formalism for many quantities, and can also be used for interpolating various physical properties on a fine mesh in the Brillouin zone~\cite{PhysRevB.74.195118,PhysRevB.76.165108}. For examples, WFs can be used to compute electronic polarization, orbital magnetization, the component of isotropic magnetoelectric coupling, and various transport properties.

However, an exponentially localized Wannier function representation does not exist for insulators with a non-zero Chern number $C$.\cite{PhysRevB.74.235111,PhysRevLett.98.046402}  Insulators with a non-zero Chern number are called integer quantum Hall insulators (or Chern insulators) and are characterized with a non-zero Hall conductance $\sigma = C e^2/h$. (Three-dimensional insulators are characterized by a triplet of Chern numbers.)

In the past several years there has been significant interest in a group of materials related to the Chern insulator.  These  are  called $\mathbb{Z}_2$ topological insulators (TIs).  In two dimensions these $\mathbb{Z}_2$ topological insulators can be seen as topologically equivalent to two copies of a Chern insulator, one with $C=1$ and another with $C=-1$,
\begin{equation}
  H_{\rm TI} = H_{+1} \oplus H_{-1}.
\label{eq:pmC}
\end{equation}
Therefore the Chern number for a $\mathbb{Z}_2$ topological insulator is zero\cite{PhysRevLett.98.046402} which guarantees that it allows exponentially localized WFs ($C$ is additive over bands so in this case we have $C=1-1=0$).

However, WFs of $\mathbb{Z}_2$ topological insulators do not preserve time reversal (TR) symmetry~\cite{PhysRevB.83.035108} even though the underlying Hamiltonian $H_{\rm TI}$ itself is time-reversal symmetric.  This can be seen by realizing that constructing TR-preserving WFs would be equivalent to constructing WFs individually for the bands given by $H_{+1}$ and $H_{-1}$ separately, which is not possible as bands with non-zero $C$ don't have exponentially localized WFs.  Therefore, the only possibility for constructing a smooth gauge of a compound system ($H_{\rm TI}$) is to break TR symmetry in the gauge by mixing eigenstates of $H_{+1}$ with those of $H_{-1}$.

\subsection{Constructing Wannier functions from a guess}\label{sec:wannier90}
Generalized WFs~\cite{MLWFs-MV} are obtained as the Fourier transform of the Bloch states $\psi_{m\kk}$ (here we consider the case of three dimensions)
\begin{equation}\label{eq:wf}
  \Ket{\RR n}=\frac{V}{\left(2\pi\right)^3}\int\text{d}\kk e^{-i\kk\cdot\RR}\sum_m u^{(\kk)}_{mn}\Ket{\psi_{m\kk}},
\end{equation}
where $u^{(\kk)}$ is an arbitrary unitary matrix that mixes different bands for a given $\kk$-point, $\RR$ is a translation vector, and $n$ is an integer running over the number of bands considered. This gauge freedom can be used to construct WFs with minimal possible spatial extent.  These so called maximally localized Wannier functions (MLWFs) minimize the spread functional $\Omega$,
\begin{equation}\label{eq:spread_r}
  \Omega = \sum_n \left(\Braket{r^2}_n-\Braket{\rr}^2_n\right),
\end{equation}
with
\begin{align}
  \Braket{r^2}_n &= \Braket{\mathbf{0}n|r^2|\mathbf{0}n}, \\
  \Braket{\rr}_n &= \Braket{\mathbf{0}n|\rr|\mathbf{0}n}.
\end{align}

Due to the properties of the Fourier transform, localization of Wannier function $\ket{\mathbf{0} n}$ in real space is equivalent to the smoothness of Bloch states $\sum_m u^{(\kk)}_{mn}\Ket{\psi_{m\kk}}$ in $\kk$-space.

Within the standard approach~\cite{MLWFs-MV} MLWFs are constructed for a set of $N$ composite bands by first guessing a set of $N$ localized orbitals $g_n(\rr)$ that are close to the $N$ target Wannier functions,
\begin{equation}\label{eq:good-guess}
  \Ket{g_n}\approx\Ket{\bm{0}n}.
\end{equation}
Now given a set of Bloch states $\Ket{\psi_{m\kk}}$ that are potentially not smooth in $\kk$-space (or equivalently its WFs are not localized) one can try smoothening them by first projecting them into these guess orbitals $g_n$
\begin{equation}
  a^{(\kk)}_{mn} = \Braket{\psi_{m\kk} | g_n}
\end{equation}
and then constructing the unitary matrices $u^{(\kk)}$ via the L{\"o}wdin orthonormalization procedure~\cite{Lowdin},
\begin{equation}\label{eq:gauge0}
  u^{(\kk)} = a^{(\kk)}\left[a^{(\kk)\dagger}a^{(\kk)}\right]^{-1/2}.
\end{equation}
If the overlap matrix appearing above under the inverse square root
\begin{equation}\label{eq:def-sk}
  s^{(\kk)} \equiv a^{(\kk)\dagger}a^{(\kk)}
\end{equation}
has large singular values then this procedure is well-defined and matrices $u^{(\kk)}$ constructed in this manner can be used to rotate Bloch states into a smooth gauge $u^{(\kk)} \Ket{\psi}$.

It is trivial to show that if $\Ket{g_n}$ are MLWFs, that singular values of $s^{(\kk)}$ all equal $1$ and this procedure gives back rotated Bloch states $u^{(\kk)} \Ket{\psi}$ that correspond to the original MLWFs.  Therefore, one can hope that if the $\Ket{g_n}$ are at least close to MLWFs the resulting rotated Bloch states again correspond to localized---but not necessarily maximally localized---WFs.  Given this starting point one can use procedure from Ref.~\onlinecite{MLWFs-MV} to arrive at MLWFs if needed.

On the other hand if the $\Ket{g_n}$ are not close to MLWFs then the resulting rotated Bloch states need not be smooth.  This is easily seen in the case of a single band.  In this case if the orbital $\Ket{g}$ fails to capture the character of the Bloch state $\Ket{\psi_{\kk}}$ for some $\kk$-point then the complex number $a^{(\kk)}$ will have a small norm at that $\kk$.  If the norm of $a^{(\kk)}$ is exactly zero for some $\kk$ then the procedure involving a negative square root in Eq.~\eqref{eq:gauge0} is ill-defined as it involves division by zero.  However, if the norm of $a^{(\kk)}$ is small but non-zero then the procedure is potentially numerically unstable as small noise in $a^{(\kk)}$ might get amplified when taking the negative square root of $s=a^{\dagger} a$. This analysis generalizes to the case of multiple occupied bands ($N > 1$) in the following way.  If one of the rotated $N$ Bloch states are not captured well by guess $\Ket{g_n}$ then the overlap matrix (which is now a full $N\!\times\!N$ matrix) will have one small singular value and again the process of taking the negative square root of $s^{(\kk)}$ in Eq.~\eqref{eq:gauge0} will be ill-defined or unstable.

In fact this is precisely the way in which the L{\"o}wdin procedure fails if one tries to apply it to the case of a Chern insulator.~\cite{PhysRevB.74.235111}  For any localized trial orbital $\Ket{g_n}$ the overlap matrix $s^{(\kk)}$ for a Chern insulator will have at least one zero singular value at least at one point in the Brillouin zone.  This will also happen in the case of a $\mathbb{Z}_2$ topological insulators if one chooses trial orbitals that form a time-reversal symmetric pair. In Ref.~\citenum{PhysRevB.83.035108} it was recognized that projecting trial orbitals that break TR symmetry is necessary to ensure that all singular values of $s^{(\kk)}$ are nonzero everywhere in the Brillouin zone. In practice this approach still requires an initial guess of orbitals that approximate the target WFs.  This guessing is somewhat harder than in the case of a non-topological insulator since it must break TR and potentially some other crystalline symmetries. In Ref.~\citenum{PhysRevB.83.035108} this was achieved for a tight-binding model by an educated guess of trial orbitals based on the orbital character of the bands at the band inversion points and symmetries present in the model.

Another approach for constructing WFs for $\mathbb{Z}_2$ topological insulators was introduced in Ref.~\citenum{PhysRevB.93.035453}.  This approach relies on constructing a smooth gauge in a closely related non-topological insulator phase and then transporting that gauge to the TI of interest by following a path in the parameter space that explicitly breaks time-reversal symmetry (and potentially other symmetries such as inversion). This parameter space has to be chosen for each system at hand by adding terms to the Hamiltonian that break TR (and potentially crystalline) symmetry and keep the electron band gap open.

In this manuscript we will present a method that can automatically construct WFs for topologically nontrivial insulators.

\section{Our approach, the OPFM}
\label{sec:opfm}

In a recent manuscript we introduced the optimized projection functions method (OPFM)\cite{OPFM} that allows automatic construction of MLWFs. We will now give a brief review of the OPFM and then discuss why this approach is suitable for constructing WFs for $\mathbb{Z}_2$ topological insulators.

As opposed to the standard approach, which requires $N$ trial orbitals for  $N$ composite electron bands, in the OPFM one selects a larger set of $M\!>\!N$ orbitals $h_i(\rr)$ that approximately span the space of $N$ Wannier functions in a home cell,
\begin{equation}\label{eq:span}
  \text{Span}(\Ket{h_i})\supseteq\text{Span}(\Ket{\bm{0}n}).
\end{equation}
This can easily be achieved by including in $\{h\}$ valence atomic orbitals.

Given a set of projection orbitals $\{h\}$, we use the OPFM to find a semiunitary $M\!\times\!N$ matrix W such that the $N$ orbitals
\begin{equation}
  \Ket{\widetilde{g}_j}=\sum_{i=1}^M W_{ij}\Ket{h_i}
\end{equation}
are as close as possible to localized WFs.  Given the functions $\widetilde{g}_j$ one can construct the smooth gauge by first expanding the original functions into Bloch states,
\begin{equation}
  A^{(\kk)}_{mn}=\Braket{\psi_{m\kk} | h_n}
\end{equation}
and then rotating them into the optimal subset,
\begin{equation}
  a^{(\kk)}_{mn} = A^{(\kk)}W
\end{equation}
which can then be used in the L{\"o}wdin procedure.

Now we discuss why the OPFM  is suitable for constructing WFs in topological insulators.   In $\mathbb{Z}_2$ topological insulators, the spin-orbit interaction induces a band-inversion between states of different orbital character.  For example, in the case of Bi$_2$Se$_3$ the topologically nontrivial state is induced by a band inversion at the $\Gamma$ point between Se and Bi states.  Therefore one can expect that the MLWFs for the occupied bands in this system will contain a mixture of both Se and Bi states.  Guessing such a mixture is nontrivial for several reasons.  First, as we will show later, this mixture includes complex imaginary components.  Second, the mixture typically contains a contribution from more than two atoms.  Third, the mixture must break all relevant symmetries which enforce the topologically nontrivial state.  However, our OPFM can find this mixture since $\{h\}$ in the case of Bi$_2$Se$_3$ can include both Se and Bi atomic orbitals as $M$, the number of orbitals in set $\{h\}$, can be larger than the number of electron bands $N$.

In this manuscript we follow the notation we introduced in Ref.~\onlinecite{OPFM} where square $N\!\times\!N$ matrices are represented by lowercase letters (e.g.\ $a^{(\kk)}$ and $u^{(\kk)}$), and larger rectangular $N\!\times\!M$, $M\!\times\!N$, or square $M\!\times\!M$ matrices are represented by uppercase letters (e.g.\ $A^{(\kk)}$ and $W$).

\subsection{Selecting the set \texorpdfstring{$\{h\}$}{h}}\label{sec:basis-h}
Now we will discuss a choice of orbitals $h_i$ that satisfy the condition given in Eq.~\eqref{eq:span}.  Mathematically speaking, without knowing anything about chemical bonding in the insulator of interest, one would have to include in set $\{h\}$ all atomic orbitals on all atoms in the crystal to guarantee a complete basis for expansion of the WFs. Luckily, in an ionic or a covalent insulator it is enough to choose set $\{h\}$ to include only  valence atomic orbitals, as they are typically the ones forming atomic bonds.  In addition, since WFs are typically exponentially localized it is enough to choose orbitals in the home cell, and possibly few atoms in the neighboring unit cells.  (For example, as discussed in Ref.~\onlinecite{OPFM} in the case of cubic silicon one has to include in set $\{h\}$ atomic orbitals centered on two atoms in the basis as well as three neighboring atoms, so that, for each of the four Si-Si bonds, both Si atoms forming a particular bond are included in the set $\{h\}$.)

However, bonding in the case of $\mathbb{Z}_2$ TIs is more involved than in a typical insulator. As we will show in this manuscript, presence of spin-orbit induced band inversion induces an intricate bonding network so that some WFs extends over more than two atoms (as in the case of the bonds in silicon) and thus one needs to use a somewhat larger set $\{h\}$ than in a conventional covalent material. However, in all cases we tested, it was enough to include in $\{h\}$ orbitals in the home-cell along with the orbitals in a single neighboring cell.

\subsection{Finding matrix \texorpdfstring{$W$}{W}}\label{sec:offdiag}
In Ref.~\citenum{OPFM} the problem of finding $W$ that minimizes the WF spread $\Omega$ was reduced to minimizing the Lagrangian
\begin{align}\label{eq:lagrangian_old}
\mathcal{L} \left(W,\lambda\right) =
  \ &\widetilde{\Omega}_\text{I,OD}(W) \\
  &+ \lambda w \sum_{\kk}\sum_{i=1}^N \abs*{\left[W^{\dagger}\left( S^{(\kk)} -I_M \right) W\right]_{ii}}^2, \notag
\end{align}
where we define the large overlap matrix
\begin{equation}
  S^{(\kk)} \equiv A^{(\kk)\dagger}A^{(\kk)}.
\end{equation}
The first term in Eq.~\eqref{eq:lagrangian_old} approximates the sum of the invariant and offdiagonal parts of the spread $\Omega$.  However, this approximation is valid only when rotated overlap matrix $W^\dagger S^{(\kk)} W$ is close to the identity matrix (see discussion in Ref.~\onlinecite{OPFM}).  For simply bonded insulators this condition is enforced by the second term in Eq.~\eqref{eq:lagrangian_old}.

While strictly speaking the entire matrix $W^\dagger S^{(\kk)} W$ should be close to the identity matrix, the second term in Eq.~\eqref{eq:lagrangian_old} only penalizes the deviation of diagonal elements of $W^\dagger S^{(\kk)} W$ away from $1$.  This simplification is adequate for the case of simply bonded  insulators  where only a small number of atoms are needed to span the space of WFs centered in the home cell.  However, for the case of $\mathbb{Z}_2$ TIs, one needs to use a somewhat larger set $\{h\}$ and this simplification is insufficient since now some contributions to a WF could potentially be duplicated by more than one element in the set $\{h\}$.  

Therefore, in this manuscript we will construct $W$ by minimizing the following Lagrangian that penalizes the offdiagonal elements of $W^\dagger S^{(\kk)} W$ as well,
\begin{align}\label{eq:lagrangian_new}
\begin{split}
\mathcal{L} \left(W,\lambda\right) =
  \ &\widetilde{\Omega}_\text{I,OD}(W) \\
  &+ \lambda w \sum_{\kk}\sum_{i=1}^N\sum_{j=1}^N \abs*{\left[W^{\dagger} \left( S^{(\kk)} - I_M \right) W\right]_{ij}}^2 .
  \end{split}
\end{align}
We describe the algorithm to minimize such Lagrangian in App.~\ref{sec:appndx-offdiag}.

We confirmed that with this extended Lagrangian WF spreads for conventional insulators investigated in Ref.~\citenum{OPFM} are unaffected. For example, the initial spread in the case of cubic silicon changes by less than 0.2\% when offdiagonal elements are included in $\mathcal{L}$. The only difference with respect to Ref.~\citenum{OPFM} is that now---with the Lagrangian from Eq.~\eqref{eq:lagrangian_new}---it doesn't matter whether some WFs can be represented by orbitals from $\{h\}$ in more than one way, at a small additional cost in the computational time.  For example, in the case of the cubic silicon if we include 6 (instead of 3) neighboring atoms in the set $\{h\}$ so that three out of four Si-Si bonds can be represented in duplicated ways, the total initial spread is changed only by 0.6\%.

We note here that it is numerically straightforward to construct an arbitrary set $\{h\}$ given a set of orbitals $h_i$ in the home cell.  For a particular orbital $h_i(\rr)$ on the basis atom, a projection onto another orbital given by the same orbital but translated by lattice vector $\RR$ is simply,
\begin{equation}\label{eq:expand-A}
  \Braket{\psi_{\nk} | h_i(\rr-\RR)} = e^{-i\kk\cdot\RR} \Braket{\psi_{\nk} | h_i(\rr)}
\end{equation}
by the virtue of Bloch's theorem.

\section{Examples}\label{sec:examples}
In the following subsections, we apply the optimized projection functions method to three examples of $\mathbb{Z}_2$ topological insulators. The first is the Kane-Mele model~\cite{PhysRevLett.95.146802}, which is a two-dimensional tight-binding model on the honeycomb structure. The second is a three-dimensional tight-binding model of a strong topological insulator that was introduced in Ref.~\onlinecite{Qi2008}. The third example is a realistic case of a three-dimensional strong topological insulator (Bi$_2$Se$_3$) as calculated within the density functional theory approach.~\cite{Xia2009,Zhang2009}

\subsection{Two-dimensional model}
The Kane-Mele model is a two-dimensional model of a $\mathbb{Z}_2$ topological insulator.  It contains four electron bands, two of which are considered to be occupied. It is defined on a honeycomb structure that can be described in terms of the hexagonal lattice with primitive lattice vectors $\mathbf{a}_{1,2} = \frac{a}{2}(\sqrt{3}\hat{\mathbf{y}}\pm\hat{\mathbf{x}})$, and a basis of two sites, $A$ and $B$, located at $\bm{\tau}_A=a\hat{\mathbf{y}}/\sqrt{3}$ and $\bm{\tau}_B=2a\hat{\mathbf{y}}/\sqrt{3}$, respectively. In what follows, we choose $a=1$~\AA\ for convenience.

The Kane-Mele Hamiltonian is
\begin{align}
  \begin{split}
  H =
  t&\sum_{\langle ij \rangle}c_i^\dagger c_j
  +  i\lambda_{\text{SO}}\sum_{\langle\langle ij \rangle\rangle}\nu_{ij}c_i^\dagger s^z c_j \\
  +& i\lambda_{\text{R}}\sum_{\langle ij \rangle}c_i^\dagger(\mathbf{s}\times\hat{\mathbf{d}}_{ij})_z c_j
  +  \lambda_v\sum_i \xi_i c_i^\dagger c_i.
  \end{split}
  \label{eq:KM}
\end{align}
We suppressed spin indices on the electron creation and annihilation operators. Symbol $\langle ij \rangle$ indicates a sum over nearest neighbors and $\langle\langle ij \rangle\rangle$ indicates a sum over next-nearest neighbors. The first term in the Hamiltonian is the nearest neighbor hopping, with hopping strength $t$ that we set equal to $1$ for convenience (i.e., all energies are in units of $t$). The second term describes spin-dependent second nearest neighbor hopping, which emulates a spin-orbit interaction. Here, $\nu_{ij}$ takes on the value $\pm1$ depending on the sign of $(\hat{\mathbf{d}}_1\times\hat{\mathbf{d}}_2)_z$,  where $\hat{\mathbf{d}}_1$ and $\hat{\mathbf{d}}_2$ are the unit vectors along the bonds traversed as the electron hops from site $j$ to $i$, and $s^z$ is the Pauli spin matrix. The third term describes nearest neighbor Rashba coupling, where $\hat{\mathbf{d}}_{ij}$ is the unit vector along the bond from $j$ to $i$. Lastly, the fourth term introduces a staggered on-site potential ($\xi_i=\pm1$) between the $A$ and $B$ sublattices; we choose $\xi_i$ so that the on-site potential is negative on the $B$ sublattice and the occupied bands in the normal phase have dominant $B$ character. In the following we set the staggered on-site term $\lambda_v=1$ and the Rashba term $\lambda_\text{R}=0.5$. Increasing the strength of the spin-orbit term $\lambda_\text{SO}$ tunes the model from describing the normal to the topological insulator phase, with the transition at $\lambda_\text{SO}\approx0.27$.  For calculations in the topological phase we use $\lambda_\text{SO}=0.6$.

\begin{figure}
  \includegraphics{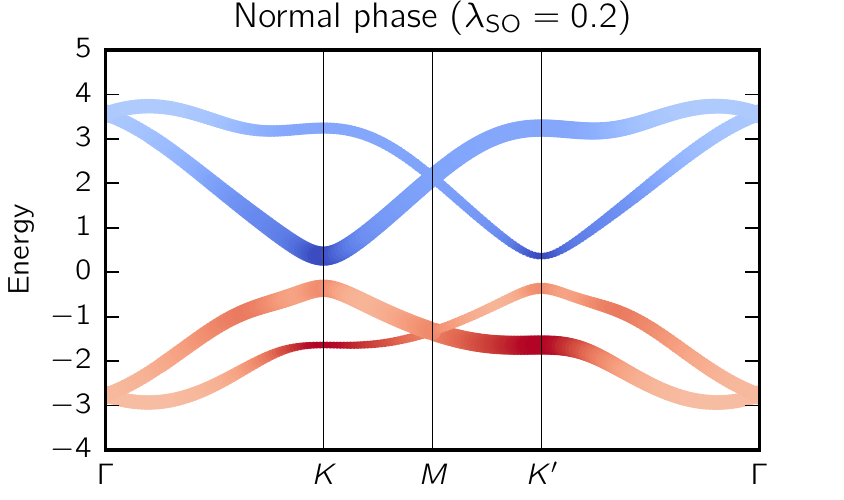}
  \par\vspace{0.25cm}
  \includegraphics{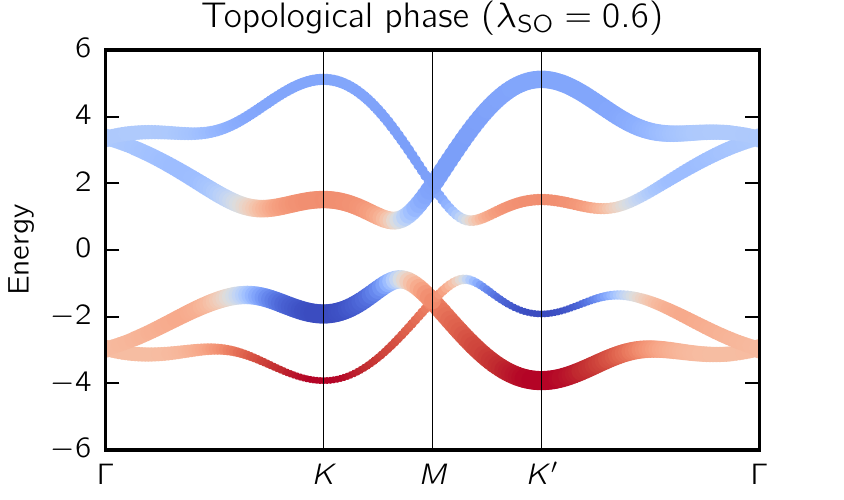}
  \caption{\label{fig:km-bands}%
  Band structures of the Kane-Mele model in the normal phase (top) and topological phase (bottom). The part of the bands colored red correspond to $B$-orbital character, while blue corresponds to $A$-orbital character. The thickness of the line corresponds to the spin component, with thicker indicating mostly spin-up, and thinner indicating spin-down. The zero of energy is set to the middle of the gap.}
\end{figure}

The Kane-Mele model is solved using the PythTB~\footnote{\url{http://physics.rutgers.edu/pythtb/}} package with a basis of two orbitals per site, one each for spin-up and spin-down,
\begin{equation}\label{eq:km-orbitals}
  \Ket{A;\uparrow_z}, \quad
  \Ket{A;\downarrow_z}, \quad
  \Ket{B;\uparrow_z}, \quad
  \Ket{B;\downarrow_z}.
\end{equation}
In Figure~\ref{fig:km-bands} we plot the band structures in both the normal phase ($\mathbb{Z}_2$ even) and the topological phase ($\mathbb{Z}_2$ odd). The bands are colored according to the character of the state, with red corresponding to a state of $B$-orbital character and blue corresponding to $A$-orbital character, and gray indicating a mixture.
The thickness of the line indicate the spin of the state along the $z$-axis, with thicker corresponding to mostly spin-up and thinner line corresponding to mostly spin-down; an intermediate thickness indicates mixed spin state due to the Rashba-like term. In the topological phase, there is a clear inversion of the character of the states near the $K$ and $K^\prime$ points.

\subsubsection{Selecting the set \texorpdfstring{$\{h\}$}{h}}\label{sec:cluster}

Our results of the OPFM applied to the case of a Kane-Mele model are shown at the top of Table~\ref{tab:km} along with the result form previous work.\cite{PhysRevB.83.035108}  In the previous work the following guess orbitals were used to construct WFs for the Kane-Mele model in the topological phase:
$\Ket{A;\uparrow_x}$~and~$\Ket{B;\downarrow_x}$. Note that spins here point in plane ($x$) while the basis functions have spins pointing perpendicular to the plane ($z$).

To construct localized WFs for the Kane-Mele model using the OPFM we consider several different sets $\{h\}$ of basis functions. The smallest set consisted of the four orbitals on the two atoms in the home cell defined in Eq.~\eqref{eq:km-orbitals}. As expected, this small set is unable to capture the extended nature of the WF in the $\mathbb{Z}_2$ topological insulator. The fact that this set is too small is numerically indicated with a small minimal singular value of $s^{(\kk)}=W^{\dagger}S^{(\kk)}W$ (its value is only 0.03, not given in Table~\ref{tab:km}) which then results in an ill-defined L{\"o}wdin orthonormalization procedure.  This procedure produces a gauge with a WF spread $\Omega^0=0.333$, significantly higher than that of a MLWF ($\Omega^{\rm GM}=0.189$). If we try smoothening this gauge further with the Marzari-Vanderbilt procedure\cite{MLWFs-MV} it remains stuck in a local minimum as the spread is only slightly reduced to $\Omega^{\rm GM}=0.319$.

Since the figure of merit (i.e. minimal singular value of $s^{(\kk)}$) was small for this set $\{h\}$ we decided to use the OPFM with larger sets $\{h\}$. The next set we considered, labeled $\{0,1^{\text{st}}\}$, includes---in addition to orbitals in the home cell---orbitals on their four first-neighboring atoms. An even larger set we tried $\{0,1^{\text{st}},2^{\text{nd}}\}$ includes both first and second nearest neighboring atoms.

\begin{table*}
\caption{\label{tab:km}%
  Results of the OPFM applied to the examples in Section~\ref{sec:examples}. The first column gives the square moduli of $s^{(\kk)}-I_N$ averaged over $\kk$-points and matrix elements. The second column lists the minimal singular value of $s^{(\kk)}$ over all $\kk$-points. The third column show the spread $\Omega^0$ after L{\"o}wdin procedure while the fourth column shows the spread $\Omega^{\rm GM}$ when  L{\"o}wdin procedure is followed up by the Marzari-Vanderbilt\cite{MLWFs-MV} procedure. For three-dimensional cases, the fifth column gives the value for the Chern-Simons $\theta$ term. See text for a description of the sets $\{h\}$ used in the OPFM.
  }
\begin{ruledtabular}
\begin{tabular}{lccccc}
 & \multirow{2}{*}{Average $\abs*{\left(s^{(\kk)} - I_N \right)_{ij} }^2$} 
 & \multirow{2}{*}{Min. $\text{sing}\left( s^{(\kk)} \right) $} 
 & \multicolumn{2}{c}{Spread (\AA$^2$)}
 & \multirow{2}{*}{Chern-Simons $\theta$} \\
\cline{4-5}
 & &  & \multicolumn{1}{c}{$\Omega^0$} & \multicolumn{1}{c}{$\Omega^\text{GM}$} & \\
\hline
\vspace{-6pt}\\
% ------------------------------------------------------------------------------
\multicolumn{6}{l}{Two-dimensional model}\\

\quad\quad Previous work, Ref.~\onlinecite{PhysRevB.83.035108}
 & 0.148 & \multicolumn{1}{d{1.5}}{0.11} & 0.212 & 0.189 &  \\

\quad\quad OPFM using set $\{ 0, 1^{\text{st}} \}$
 & 0.017 & \multicolumn{1}{d{1.5}}{0.40} & 0.244 & 0.189 &  \\

\quad\quad OPFM using set $\{ 0, 1^{\text{st}}, 2^{\text{nd}} \}$
 & 0.006 & \multicolumn{1}{d{1.5}}{0.71} & 0.207 & 0.189 &  \\

% ------------------------------------------------------------------------------
\vspace{-6pt}\\
\multicolumn{6}{l}{Three-dimensional model}\\

\quad\quad OPFM using set $\{ 0, 1 \}$      
 & 0.0184 & \multicolumn{1}{d{1.5}}{0.0472} & 0.142 & 0.135 & 0.96$\pi$       \\

\quad\quad OPFM using set $\{ 0, 1, 2, 3 \}$
 & 0.0133 & \multicolumn{1}{d{1.5}}{0.1022} & 0.142 & 0.135 & 0.96$\pi$       \\

% ------------------------------------------------------------------------------
\vspace{-6pt}\\
\multicolumn{6}{l}{Density functional theory, Bi$_2$Se$_3$}\\

\quad\quad Previous work, Ref.~\onlinecite{Coh-CSOMP}
 & 0.0068 & \multicolumn{1}{d{1.5}}{0.0003}  & 109.80 & 95.84 & 0.32$\pi$ \\

\quad\quad Previous work, Ref.~\onlinecite{PhysRevB.93.035453}
 & 0.0057 & \multicolumn{1}{d{1.5}}{0.0002}  & 126.12 & 95.85 & 0.35$\pi$ \\

\quad\quad OPFM using set $\{ 0,1 \}$  
 & 0.0026 & \multicolumn{1}{d{1.5}}{0.0001} & 133.43 & 95.84 & 0.34$\pi$ \\

\quad\quad OPFM using set $\{ 0,1,2,3 \}$  
 & 0.0017 & \multicolumn{1}{d{1.5}}{0.0240}  & 310.29 & 95.83 & 0.34$\pi$ \\

% ------------------------------------------------------------------------------
\end{tabular}
\end{ruledtabular}
\end{table*}

As soon as we include first or second neighbor atoms into the set $\{h\}$ the resulting minimal singular value of $s^{(\kk)}$ increases from 0.03 to 0.40 and 0.71, respectively for the two sets, and the resulting spread $\Omega^0$ decreases.  Final spread $\Omega^{\rm GM}$ agrees with previous result\cite{PhysRevB.83.035108} up to numerical precision.

\begin{figure}
  \includegraphics[width=\columnwidth]{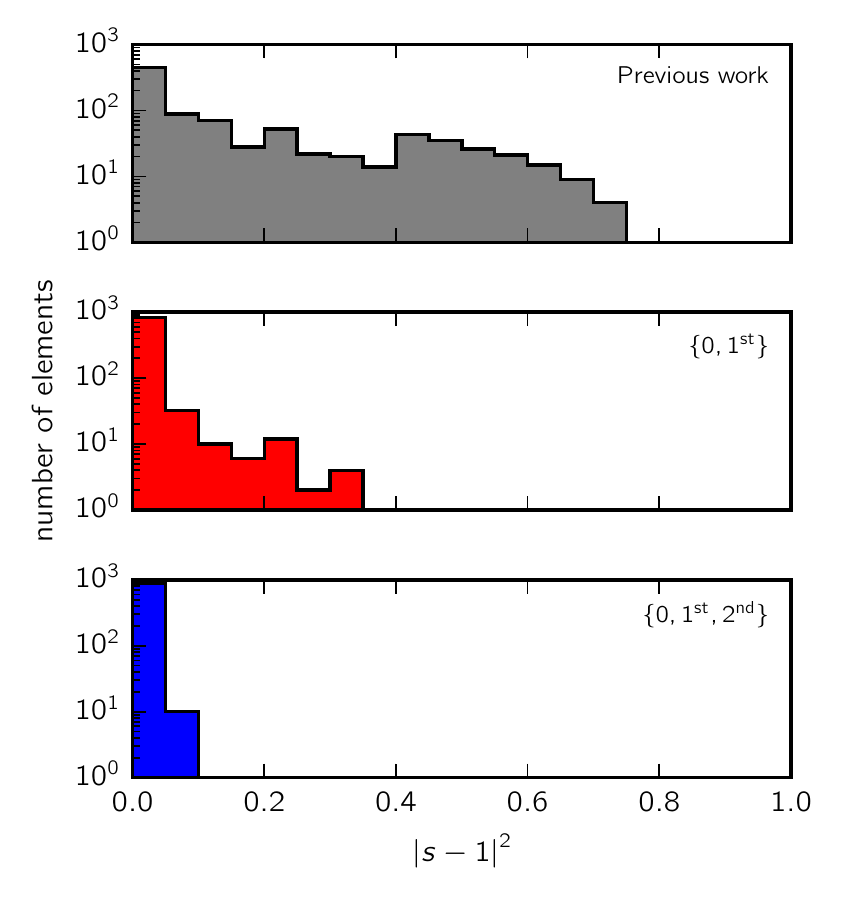}
  \caption{\label{fig:km-overlap-cluster}%
  Histograms of the square moduli of the elements of $s^{(\kk)}-I$ for guess orbitals from Ref.~\citenum{PhysRevB.83.035108} (top panel) and two sets $\{h\}$ within the OPFM (middle and bottom panel).
  }
\end{figure}

Table~\ref{tab:km} contains also the average distance between the overlap matrix $s^{(\kk)}$ and the identity matrix.  However, that averaged quantity masks the fact that the overlap matrix is typically different from an identity matrix only in a small part of the Brillouin zone where inversion occurs (see band characters near $K$ and $K'$ points in Fig.~\ref{fig:km-bands}).  Therefore for the purpose of presentation we give in Fig.~\ref{fig:km-overlap-cluster} the distribution of $\abs*{\left(s^{(\kk)}-I\right)_{ij}}^2$ over all $\kk$-points and all its matrix elements $ij$. Note that this quantity is the same as the second term in Eq.~\eqref{eq:lagrangian_new}.

As can be seen from Fig.~\ref{fig:km-overlap-cluster} in all cases  distance of $s^{(\kk)}$ from identity is small for nearly all $\kk$-points (note that the vertical scale is logarithmic).  However, in the case of the guess orbitals from  Ref.~\onlinecite{PhysRevB.83.035108} there is a fraction of $\kk$ points for which matrix elements of $s^{(\kk)}$ are quite far away from identity matrix (up to 0.7).  These $\kk$-points correspond to the small part of the Brillouin zone with inverted bands. However, singular values of $s^{(\kk)}$ are large enough (smallest one is 0.11) so that the L{\"o}wdin procedure is well behaved even for the guess orbital from Ref.~\onlinecite{PhysRevB.83.035108}.  In the case of the OPFM the deviation of $s^{(\kk)}$ from identity is significantly smaller (the maximum value is only 0.1 for the cluster $\{ 0, 1^{\text{st}}, 2^{\text{nd}} \}$).

We also confirmed that OPFM gives automatically good spread as one varies $\lambda_\text{SO}$ through the transition from the topological all the way to the normal phase.

For this two-dimensional model we constructed the projection matrices on a $15^2$ $\kk$-point grid. We find that the optimal value for the Lagrange multiplier ($\lambda$) is 0.03. We also note that while in the case of normal insulators studied previously one can often initialize $W$ with the identity matrix, in the case of $\mathbb{Z}_2$ TIs we sometimes need to start off the minimization from a random matrix so that the initial $W$ breaks all symmetries. (We confirmed that, in the case of normal insulators studied in  Ref.~\citenum{OPFM}, starting minimization procedure from a random matrix does not affect the final spread.  For example, in the case of cubic silicon the total spread is unaffected within numerical precision.)

\subsubsection{Analysis of the WF}

\begin{figure*}
  \includegraphics{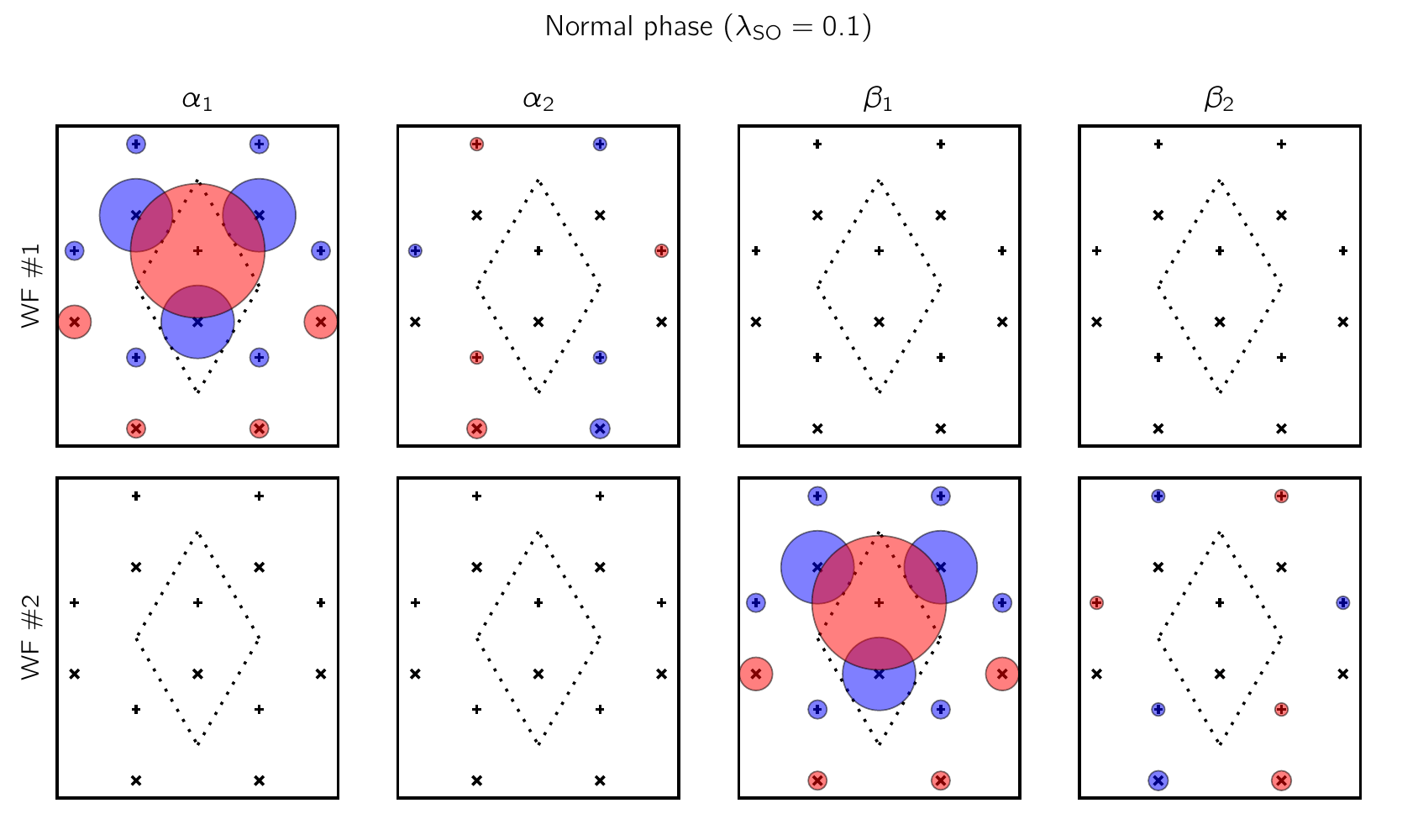}
  \par\vspace{0.75cm}
  \includegraphics{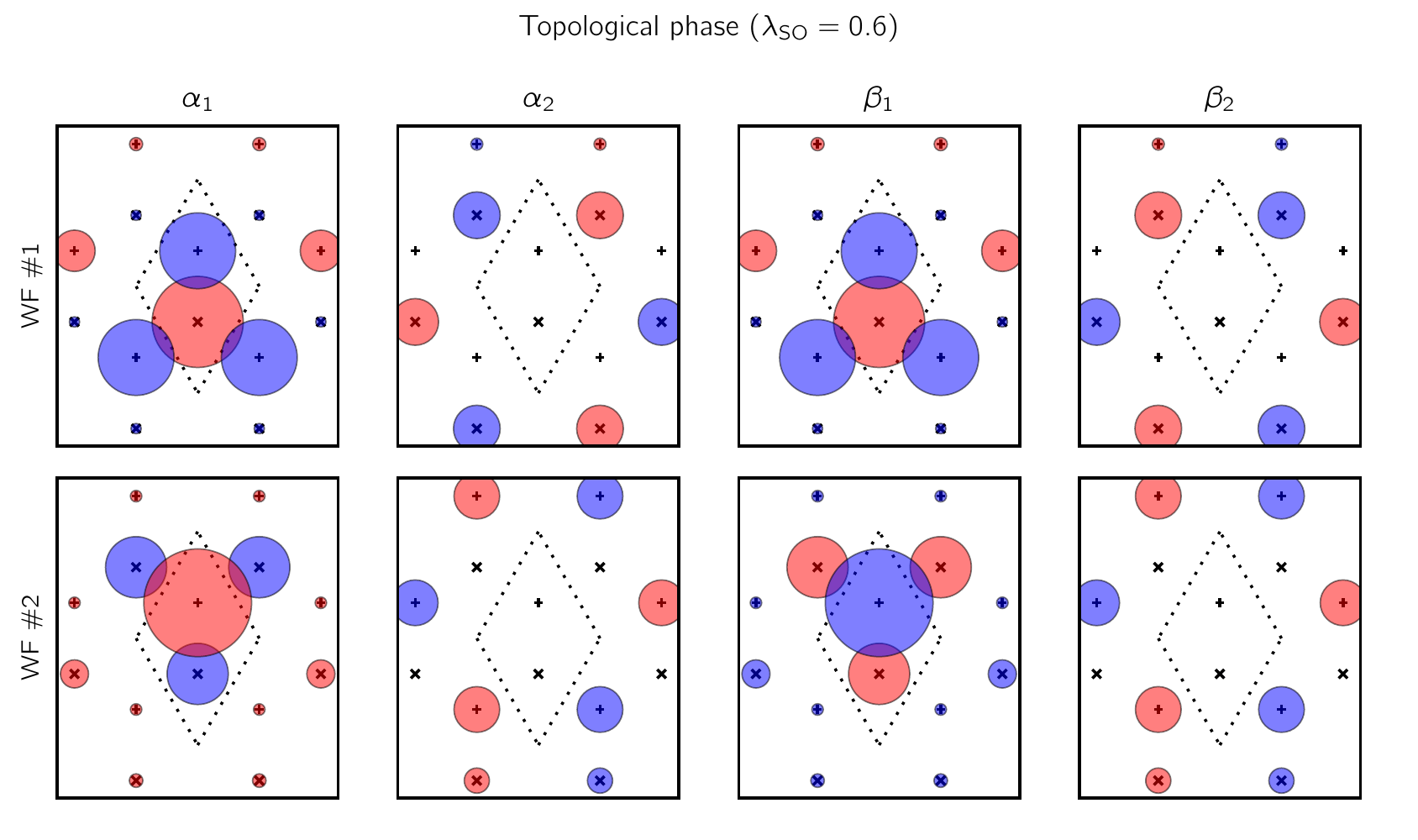}
  \caption{\label{fig:km-wfs}%
Plot of the Wannier functions obtained via the OPFM on set $\{ 0, 1^{\text{st}}, 2^{\text{nd}} \}$, for the two occupied bands of the Kane-Mele model in the normal insulator phase (top) and the topological insulator phase (bottom). The crosses ($\bm \times$) indicate the $A$ sites, while the plus signs ($\bm +$) indicate the $B$ sites. Red circles correspond to a component that is positive and blue circles correspond to negative. The size (area) of the circle is proportional to the magnitude of the component $\alpha_1,\alpha_2,\beta_1,\beta_2$. Here, the WFs plotted are for a Kane-Mele model with $\lambda_v=1$ and $\lambda_\text{R}=0$.
  }
\end{figure*}

As expected, we find that the WF in the $\mathbb{Z}_2$ topological insulator case extends well beyond the home cell.  This finding is expected since band inversion usually occurs over a small region in $\kk$-space.  To analyze the extent of the Wannier function in more detail we show in Fig.~\ref{fig:km-wfs} the MLWFs for the Kane-Mele model in the real space.  
We write the WF amplitude on a particular sublattice $j=\{A,B;\RR\}$ in the crystal as,
\begin{equation}
  \left(\alpha_1 + i\alpha_2\right)\Ket{j\uparrow} + \left(\beta_1 + i\beta_2\right)\Ket{j\downarrow}.
\end{equation}
These amplitudes can be computed from the projections of Bloch states into basis functions ($A^{(\kk)}$) and the smooth gauge for the WFs $u^{(\kk)}$ as,
\begin{align*}
\Braket{h_j | \RR n}
  &= \bra{h_j}\sum_{m\kk}u^{(\kk)}_{mn}\Ket{\psi_{m\kk}}
   = \sum_{\kk}A^{(\kk)\dagger}u^{(\kk)},
\end{align*}

Figure~\ref{fig:km-wfs} shows the amplitudes $\alpha_1$, $\alpha_2$, $\beta_1$, and $\beta_2$ on each site $j$ in the crystal for both occupied bands (labelled \#1 and \#2). The size of the circles are proportional to the absolute value of the magnitude of $\alpha_1,\alpha_2,\beta_1,\beta_2$ while color denotes their sign (red for positive and blue for negative).   The cross symbols ($\bm\times$) denote $A$ sites while plus symbols ($\bm+$) denote $B$ sites.
  
In the normal phase, both WFs are centered near the $B$ site in the home cell, with components of opposite sign on the first nearest neighbors, and small components on second nearest neighbors.

The topological phase has WFs that are centered near different sites ($A$ and $B$ in the home cell), with both being a mixture of spin-up and spin-down.  Most importantly, in the topological phase the WF amplitude extends well beyond the home cell into the first and second nearest neighbors.  The beyond-home-cell component of the WF in addition has a significant imaginary part. 

Therefore, from here we confirm once again that the set $\{h\}$ in the case of a $\mathbb{Z}_2$ topological insulator must include orbitals beyond those in the home cell. 

\subsection{Three-dimensional model}\label{sec:3d-ti}

We now turn to the model of a three-dimensional strong $\mathbb{Z}_2$ topological insulator. A simple model of such an insulator is given in Ref.~\citenum{Qi2008} by constructing a higher dimensional insulator with a non-zero second Chern number and then restricting the model to three dimensions. Similarly as in the case of a Kane-Mele model this model consists of four orbitals in the basis and two occupied electronic bands. One difference with respect to the Kane-Mele model is that the only hopping terms in the model are either between orbitals in the same unit cell, or between the first neighboring cells. The Kane-Mele model includes hopping to second nearest neighboring cells as well.

As is done for the Kane-Mele model, here we performed OPFM with sets $\{h\}$ of increasing size.  Once again we find that as soon as a neighboring cell is included in the set $\{h\}$, the OPFM procedure produces a smooth gauge.  Similarly, as in the case of Kane-Mele model, we find that with larger sets $\{h\}$ the minimal singular value of the overlap matrix is increased.  Here we adopt a notation by which the set $\{0, 1\}$ corresponds to orbitals in the home cell as well as neighboring cells translated along the first lattice vector.  Similarly, the set $\{0, 1, 2, 3\}$  corresponds to orbitals in the home cell as well as those translated along all three lattice vectors.

In addition to the quantities reported in Table~\ref{tab:km} for the Kane-Mele model, here we also report the value of the Chern-Simons orbital magnetoelectric coupling $\theta$.~\cite{Qi2008,PhysRevLett.102.146805} The $\theta$ term takes on the value 0 or $\pi$ (modulo $2\pi$) in the normal and topological phase, respectively.  However, these values would be obtained only in the limit of infinitely dense $\kk$-meshes, as the discretized expression for $\theta$ we used is not gauge invariant (gauge invariant discrete form of $\theta$ is unknown, as far as we are aware).  On a finite mesh the calculated value of $\theta$ in the topological phase is typically smaller than $\pi$ and it converges very slowly to $\pi$ as the $\kk$-mesh gets denser. We used a $20^3$ mesh of $\kk$ points for this calculation.  Here the value of the Lagrange multiplier ($\lambda$) is $1$.

The expressions for $\theta$ in terms of WFs given in Ref.~\onlinecite{Coh-CSOMP} clearly shows that $\theta$ must be $0$ if the WFs are purely real.  Therefore, just as in the case of two-dimensional model, the WFs in the three-dimensional topological insulator must contain large imaginary components so that $\theta$ can be non-zero ($\theta=\pi$ modulo $2\pi$ to be precise).

Table~\ref{tab:km} contains some of the results of the OPFM applied to the three-dimensional model. In the normal phase of that model (not shown in the Table~\ref{tab:km}) our OPFM finds projection functions that well approximate the WFs even when we use the set $\{0\}$ with orbitals only in the home cell.  The minimal singular value of the $s^{(\kk)}$ matrix is close to identity (0.92) and the spread after the L{\"o}wdin procedure agrees with the spread at the global minimum within the first four non-zero significant digits.

In the topological phase of the model, the set $\{0\}$ results in an overlap matrix with a very small singular value ($10^{-29}$) but with an inclusion of a larger set $\{h\}$ all figures of merit improve, as in the case of the Kane-Mele model.  Therefore we conclude that even in the case of three-dimensional models the WF in a $\mathbb{Z}_2$ topological insulator extends well beyond the home cell.

We note here that for relatively small sets such as $\{0,1\}$ the minimal singular value is rather small (0.0472); however, the resulting final spread is very close to the spread at the global minimum and the value of $\theta$ is close to $\pi$.  The minimal singular value increases to 0.1022 in the set $\{0,1,2,3\}$.  We also tried using an even larger set where orbitals are translated both in positive and negative direction of the lattice vector $\{0,1,\bar{1},2,\bar{2},3,\bar{3}\}$ and we find that the minimal singular value increases to 0.514.  Despite having a larger minimal singular value, $\theta$ and $\Omega^{\rm GM}$ computed from this set are up to numerical precision equal to those obtained using a much smaller set, $\{0,1\}$. Therefore we conclude that the set $\{0,1\}$ is adequate for this system even though it yields a somewhat small minimal singular value of the overlap matrix (0.0472).

To further test our method we generalized the tight-binding model from Ref.~\citenum{Qi2008} to higher number of bands.  The model from Ref.~\citenum{Qi2008} was constructed from $2n$-dimensional Clifford algebra where $n=2$.  If we use $n=3$ or $n=4$ algebras and again perform dimensional reduction to three dimensions, the resulting tight-binding model will have $2^{n-1}$ occupied bands out of $2^n$ bands.  This means that in the $n=3$ case we have $8$-band model with $4$ occupied bands, while with $n=4$ we have $16$-band model with $8$ occupied bands. Applying the OPFM to these models using the set $\{0,1,\bar{1},2,\bar{2},3,\bar{3}\}$ again produces a smooth gauge.  While the minimal singular value in the $4$-band model $(n=2$) discussed earlier is 0.514, with $8$-band model ($n=3)$ it is 0.356, and with $16$-band model ($n=4$) it is 0.360.

\subsection{Density functional theory, \texorpdfstring{B\lowercase{i}$_{2}$}{Bi2}\texorpdfstring{S\lowercase{e}$_{3}$}{Se3}}\label{sec:bi2se3}

\begin{figure}
  \includegraphics{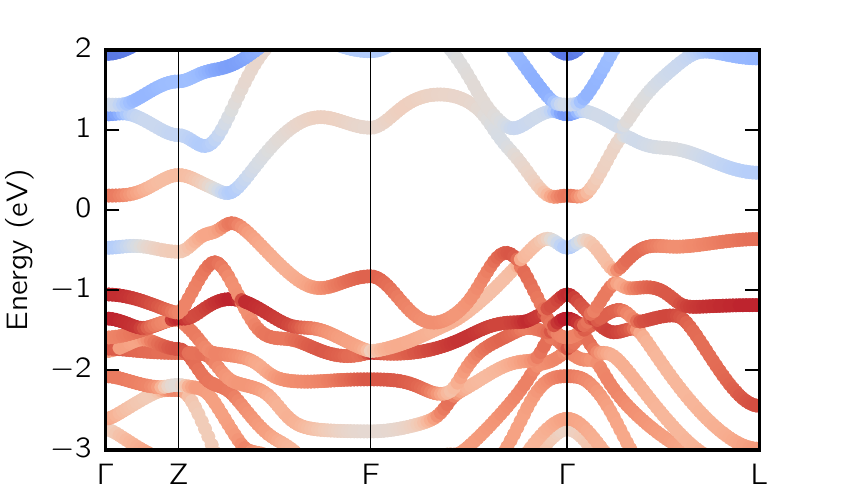}
  \caption{\label{fig:bi2se3-bands}%
  Band structure of Bi$_2$Se$_3$.
  The part of the bands colored red correspond to Se $p$ character, while blue corresponds to Bi $p$ character.
  The zero of energy is set to the middle of the gap.
  }
\end{figure}

We now turn from the tight-binding models to some realistic calculations based on density functional theory.  As an example of a prototypical strong 3D TI we use Bi$_2$Se$_3$.~\cite{Xia2009,Zhang2009} Its crystal structure is described by a rhombohedral lattice within the $D^5_{3d}$ space group. The material is made up of units of quintuple layers of Bi and Se.  Each of the five layers in the quintuple forms a hexagonal sheet in plane. The topological phase is realized due to the strong spin-orbit coupling causing a band inversion of Se $p$ and Bi $p$ character around the $\Gamma$ point.~\cite{Zhang2009} This inversion is evident in Fig.~\ref{fig:bi2se3-bands}.

To construct localized WFs for Bi$_2$Se$_3$, we first perform fully relativistic density functional theory calculations with the {\sc Quantum ESPRESSO} package.~\cite{QE-2009} The ground state properties are obtained using a $6^3$ $\kk$-point grid and a kinetic energy cutoff of 60~Ry. The projection matrices $A^{(\kk)}_{mn}$ are obtained on a $12^3$ $\kk$-point grid by projecting the top 28 valence bands into atomic Bi and Se $s$ and $p$ orbitals.  We use Eq.~\eqref{eq:expand-A} to construct projections into orbitals translated by a lattice vector. In all calculations for Bi$_2$Se$_3$ we used the value $\lambda=1$ for the Lagrange multiplier.

Again we consider several sets $\{h\}$ generated by translating the basis atoms by different lattice vectors, as using only orbitals in the home cell once again gave very small ($10^{-6}$) minimal singular value of $s^{(\kk)}$.  As soon as a neighboring cells are included in the set $\{h\}$ the minimal singular values increase as well as the spread $\Omega^{0}$.  We used the same translation vectors as in the three-dimensional model case: $\{0,1\}$ and $\{0,1,2,3\}$.  Here again $0$ represents orbitals in the home cell while non-zero integers $1,2,3$ represent translations along the three equivalent rhombohedral lattice vectors.  For completeness, we note that we chose as basis atoms those for which the reduced coordinates in the rhombohedral frame are as small as possible (between $-1/2$ and $1/2$).

The results of OPFM in the case of Bi$_2$Se$_3$ are shown in Table~\ref{tab:km} along with the results from previous work.  One of the previous works\cite{Coh-CSOMP} guessed WFs by trying out various initial projections that break symmetries while the other\cite{PhysRevB.93.035453} found it by constructing a path in parameter space that breaks time-reversal and inversion symmetry. In both earlier works the Bloch states were projected into hydrogen-like orbitals.  We find very good agreement in both $\theta$ and $\Omega^{\rm GM}$ between our approach and two earlier works. The computed value for $\theta$ in all three cases is close to $\theta\approx 0.3\pi$ since we used a relatively small $\kk$-point grid (it was $12^3$).  With larger $\kk$-grids $\theta$ converges towards $\pi$.

We note here that the minimal singular value of $s^{(\kk)}$ using relatively large set $\{ 0,1,2,3 \}$ in OPFM is still somewhat small (0.0240) even though it is two orders of magnitude larger than those in previous works.\cite{Coh-CSOMP,PhysRevB.93.035453} Nevertheless $\Omega^{\rm GM}$ agrees well with each other in all cases and the value of $\theta$ is what is expected for a three-dimensional strong $\mathbb{Z}_2$ topological insulator.  (Some of the difference between the minimal singular values in these approaches might originate from use of hydrogen-like projection functions in Refs.~\onlinecite{Coh-CSOMP,PhysRevB.93.035453})

\section{Outlook}

In this paper we described a procedure for automated construction of maximally localized Wannier functions for topologically nontrivial set of bands.  We expect that this method can be applied to any topological insulator, either protected by time-reversal symmetry, or by crystalline symmetry, as long as there exists a localized representation, i.e. as long as the first Chern numbers are all zero.  Similarly, we expect that this method could be applied to topologically nontrivial bands not only of electrons, but also of phonons, photons, cold atoms, or other particles.

\begin{acknowledgments}
We thank Bradford A. Barker for providing the pseudopotentials for Bi and Se. We also thank Georg W. Winkler for clarifying the initial projections used in Ref.~\citenum{PhysRevB.93.035453}.

This research was supported by the Theory Program at the Lawrence Berkeley National Lab through the Office of Basic Energy Sciences, U.S. Department of Energy under Contract No. DE-AC02-05CH11231 which provided the tight binding and DFT calculations;
and by the National Science Foundation under grant DMR15-1508412 which provided for basic theory and method development.
Computational resources have been provided by the DOE at Lawrence Berkeley National Laboratory's NERSC facility.
\end{acknowledgments}

\appendix

\section{Offdiagonal components}\label{sec:appndx-offdiag}

In our earlier manuscript (see  Appendix~B in Ref.~\onlinecite{OPFM}) we described an implementation of the optimized projection functions method. The appendix in the present manuscript describes how to incorporate the offdiagonals of $W^\dagger \left( S^{(\kk)} -I_M \right) W$ into Lagrangian.

\begin{figure}
  \includegraphics[width=\columnwidth]{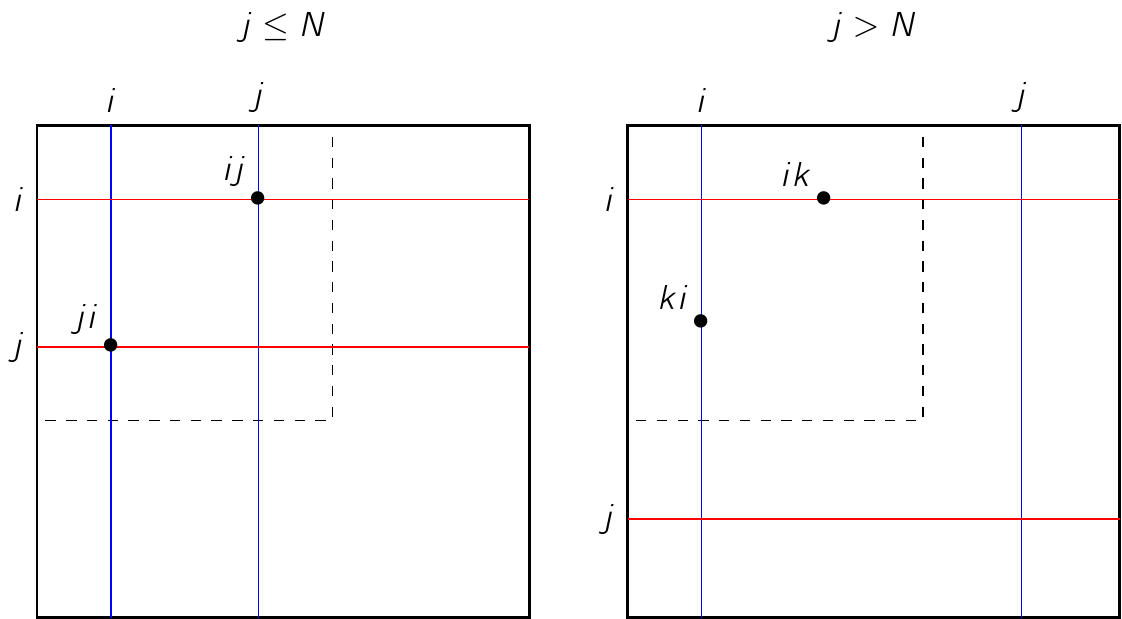}
  \caption{\label{fig:offdiag}%
  The red lines indicate the rows modified by left-multiplication by $R^\dagger$. The blue lines indicate the columns modified by right-multiplication by $R$.
  }
\end{figure}

As before, we construct the semiunitary $W$ as the $M\!\times\!N$ submatrix of a square $M\!\times\!M$ unitary matrix $\widetilde{W}$. The matrix $\widetilde{W}$ is written as a product (post-multiplication) of Givens rotations,
\begin{equation}
  \widetilde{W} = \prod_{l=1}^{L}\prod_{i=1}^{N}\prod_{j=i+1}^{M} R_{l}[i,j,\theta,\phi].
\end{equation}
A Givens rotation $R[i,j,\theta,\phi]$ ($R^\dagger[i,j,\theta,\phi]$) is a unitary planar rotation that only acts on the $i$th and $j$th columns (rows) of a matrix (see Figure~\ref{fig:offdiag}). The matrix $R[i,j,\theta,\phi]$ is identity except the $ii$, $ij$, $ji$, and $jj$ elements,
\begin{equation}
\begin{pmatrix}
R_{ii} & R_{ij} \\
R_{ji} & R_{jj}
\end{pmatrix}
=
\begin{pmatrix}
\cos\theta & e^{i\phi} \sin\theta \\
-e^{-i\phi} \sin\theta & \cos\theta
\end{pmatrix}.
\end{equation}

Again we consider two cases (see Figure~\ref{fig:offdiag}). If $j \leq N$, the $ij$ and $ji$ elements enter the Lagrangian $\mathcal{L}$. Just like the $ii$ and $jj$ components of the transformed matrices can be written in a quadratic form, so too can the offdiagonal $ij$ and $ji$ components
\begin{align}\label{eq:offdiag-ijji}
    \sum_{\alpha}& t^{(\alpha)} \left(\left\lvert\left[R^{\dagger}X^{(\alpha)}R\right]_{ij}\right\rvert^{2} + \left\lvert\left[R^{\dagger}X^{(\alpha)}R\right]_{ji}\right\rvert^{2}\right) \notag \\
      &= \mathbf{x}^{\intercal}Q\mathbf{x} + c,
\end{align}
where
\begin{equation}\label{eq:appndx-x}
  \mathbf{x}^{\intercal}=\left(\cos 2\theta,\sin 2\theta\cos\phi,\sin 2\theta\sin\phi\right).
\end{equation}
The matrix $Q$ is symmetric, and its independent components are:
\begin{align}
    2 Q_{11} =& \abs{X^{(\alpha)}_{ij}}^2 + \abs{X^{(\alpha)}_{ji}}^2 \notag \\
    2 Q_{22} =& \abs*{X^{(\alpha)}_{ii}}^2 + \abs*{X^{(\alpha)}_{jj}}^2  - 2\Re\left[X^{(\alpha)}_{ij}X^{(\alpha)*}_{ji}+X^{(\alpha)}_{ii}X^{(\alpha)*}_{jj}\right] \notag \\
    2 Q_{33} =& \abs*{X^{(\alpha)}_{ii}}^2 + \abs*{X^{(\alpha)}_{jj}}^2 + 2\Re\left[X^{(\alpha)}_{ij}X^{(\alpha)*}_{ji}-X^{(\alpha)}_{ii}X^{(\alpha)*}_{jj}\right] \notag \\
    2 Q_{12} =& \Re\left[\left(X^{(\alpha)}_{ii}-X^{(\alpha)}_{jj}\right)\left(X^{(\alpha)*}_{ij}+X^{(\alpha)*}_{ji}\right)\right] \notag \\
    2 Q_{13} =& \Re\Bigl[\left(X^{(\alpha)}_{ii}-X^{(\alpha)}_{jj}\right)\Im\left(X^{(\alpha)}_{ij}-X^{(\alpha)}_{ji}\right) \notag  \\
                & +\left(-X^{(\alpha)}_{ij}+X^{(\alpha)}_{ji}\right)\Im\left(X^{(\alpha)}_{ii}-X^{(\alpha)}_{jj}\right)\Bigr] \notag \\
    2 Q_{23} =& 2\Im X^{(\alpha)}_{ji} \Re X^{(\alpha)}_{ij} - 2\Im X^{(\alpha)}_{ij} \Re X^{(\alpha)}_{ji}.
\end{align}
The term $c$ can be ignored as it does not depend on $\bm x$. For the other offdiagonal matrix elements ($ik$, $ki$, $jk$, and $kj$, with $k \neq i \neq j \neq k$), the sum
\begin{equation}
  \abs*{X^{(\alpha)}_{ik}}^2 + \abs*{X^{(\alpha)}_{ki}}^2 + \abs*{X^{(\alpha)}_{jk}}^2 + \abs*{X^{(\alpha)}_{kj}}^2
\end{equation}
is conserved and therefore does not affect the variation of the Lagrangian.

In the case of $j>N$, the $ij$ and $ji$ elements are outside of the $N \times N$ submatrix and they therefore do not enter the Lagrangian. However, in this case when $j>N$, the sum of the square moduli of the $ik$ and $ki$ offdiagonal elements,
\begin{align}
\label{eq:offdiag-ikki}
    \sum_{\alpha}& t^{(\alpha)} \left(\left\lvert\left[R^{\dagger}X^{(\alpha)}R\right]_{ik}\right\rvert^{2}+\left\lvert\left[R^{\dagger}X^{(\alpha)}R\right]_{ki}\right\rvert^{2}\right) = \notag \\
      &= \mathbf{p}^{\intercal}\mathbf{x} + c,
\end{align}
is not conserved and it therefore must be included in minimization of $\mathcal{L}$.
The coefficient $\mathbf{p}$ of the term linear in $\mathbf{x}$ can be expressed as
\begin{align}
    p_1 =& \sum_\alpha \frac{1}{2}\left(\abs*{X^{(\alpha)}_{ik}}^2-\abs*{X^{(\alpha)}_{jk}}^2+\abs*{X^{(\alpha)}_{ki}}^2-\abs*{X^{(\alpha)}_{kj}}^2\right) \notag  \\
    p_2 =& \sum_\alpha -\Re\left[X^{(\alpha)}_{ik}X^{(\alpha)*}_{jk}+X^{(\alpha)}_{ki}X^{(\alpha)*}_{kj}\right] \notag \\
    p_3 =& \sum_\alpha \Biggl(\Im X^{(\alpha)}_{jk} \Re X^{(\alpha)}_{ik} - \Im X^{(\alpha)}_{ik} \Re X^{(\alpha)}_{jk} \notag \\
          & - \Im X^{(\alpha)}_{kj} \Re X^{(\alpha)}_{ki} + \Im X^{(\alpha)}_{ki} \Re X^{(\alpha)}_{kj}\Biggr)
\end{align}
For each $k$ (such that $k \neq i \neq j \neq k$) we have a term as in Eq.~\eqref{eq:offdiag-ikki}, with $\mathbf{p}=\mathbf{p}^{(k)}$, so that in $\mathcal{L}$ we include the term $\sum_k\mathbf{p}^{(k)\intercal}\mathbf{x}$. The terms in Eq.~\eqref{eq:offdiag-ijji} and Eq.~\eqref{eq:offdiag-ikki} are easily added to the minimization algorithm described in Ref.~\citenum{OPFM}.

\bibliography{paper}
\bibliographystyle{apsrev4-1}

\end{document}